\documentclass[aps,prl,reprint,twocolumn,superscriptaddress,nofootinbib]{revtex4}

\usepackage{amsmath,amssymb,amsthm,mathtools,bm}
\usepackage{microtype}
\usepackage{hyperref}
\usepackage{xcolor}

\hypersetup{colorlinks=true,citecolor=blue!55!black,linkcolor=blue!55!black,urlcolor=blue!55!black,pdftitle={Degeneracy Cannot Violate the Quantum Hamming Bound}}

\newtheorem{theorem}{Theorem}

\newtheorem{lemma}[theorem]{Lemma}

\newcommand{\wt}{\operatorname{wt}}
\newcommand{\Ffour}{\mathbb{F}_4}
\newcommand{\Vol}{V}
\newcommand{\Lloyd}{L}
\newcommand{\Ball}{B}

\newcommand{\one}{\mathbf{1}}
\newcommand{\doilink}[2]{\href{https://doi.org/#1}{#2}}

\begin{document}

\title{Degeneracy Cannot Violate the Quantum Hamming Bound}

\author{Yu-Xuan Zhang}
\affiliation{School of Physics, Nankai University, Tianjin 300071, People's Republic of China}

\author{Jing-Ling Chen}
\email{chenjl@nankai.edu.cn}
\affiliation{Theoretical Physics Division, Chern Institute of Mathematics, Nankai University, Tianjin 300071, People's Republic of China}

\date{June 12, 2026}

\begin{abstract}
The quantum Hamming bound is the standard finite-length sphere-packing bound for exact correction of arbitrary qubit errors. Whether degeneracy can evade this bound has remained unresolved in full generality for nearly three decades: distinct correctable errors may act identically on the code space, so the usual disjoint-sphere argument breaks down. We prove that every exact binary quantum subspace code with $K>1$ obeys the bound, without assuming either nondegeneracy or additivity. Our proof turns the Li--Xing linear-programming polynomial into an exact intersection count for quaternary Hamming balls. Monotonicity in block length and in ball-center separation then reduces the problem to a local node--edge charging inequality at the shortest admissible length. Thus degeneracy can merge correctable error sectors, but cannot enlarge the finite-length binary Hamming bound.
\end{abstract}

\maketitle

The quantum Hamming bound is the basic finite-length sphere-packing constraint on exact quantum error correction \cite{Shor1995,CalderbankShor1996,Steane1996,EkertMacchiavello1996,Laflamme1996}. For an exact binary $((n,K,d))$ quantum subspace code, let $t=\lfloor(d-1)/2\rfloor$ be the maximum number of arbitrary qubit errors that can be corrected. The bound asserts
\begin{equation}
 \Vol_t(n):=\sum_{j=0}^{t}3^j\binom{n}{j},
 \qquad K\Vol_t(n)\le 2^n.
 \label{eq:qhb}
\end{equation}
Unlike an asymptotic rate bound, Eq.~\eqref{eq:qhb} constrains every block length and every correctable radius separately. It is therefore the natural finite-length benchmark for how much logical Hilbert space can be protected against all errors affecting at most $t$ qubits. For nondegenerate codes it is the direct quantum analogue of classical sphere packing. If $P$ projects onto the code space and $E_a$ ranges over Pauli errors of weight at most $t$, exact correction is equivalent to $PE_a^\dagger E_bP=\lambda_{ab}P$ \cite{KnillLaflamme1997}. After diagonalizing the Gram matrix $(\lambda_{ab})$, nondegeneracy yields $\Vol_t(n)$ mutually orthogonal $K$-dimensional error sectors, and Eq.~\eqref{eq:qhb} follows by dimension counting. Quantum Hamming codes show that this packing picture can be attained \cite{Gottesman1996}.

Degeneracy removes exactly the step on which this argument rests. Distinct physical errors can act identically on the code space, lowering the rank of $(\lambda_{ab})$ and allowing the corresponding error sectors to overlap. The naive sphere count then overestimates the number of independent error images, with no evident substitute. This is not merely a technical pathology: degeneracy is a genuinely quantum feature that can be useful in code construction. It was therefore conceivable that sufficiently structured overlap could protect a larger subspace than nondegenerate packing permits.

Whether such overlap can beat the quantum Hamming bound in full generality has remained open since the early years of quantum error correction. By 2007 the problem had already been open for a decade \cite{Aly2007}, and it continued to be posed explicitly in later work \cite{LiXing2009,SarvepalliKlappenecker2010,Dallas2022}. Stabilizer methods alone cannot settle the question. Although the additive description over $\Ffour$ is fundamental \cite{CalderbankRainsShorSloane1998}, nonadditive codes can outperform additive constructions for some parameters \cite{RainsHardinShorSloane1997,SmolinSmithWehner2007}; a complete theorem must therefore apply to arbitrary quantum subspaces.

Successive partial results have narrowed the possible region of violation. Quantum weight enumerators and linear programming produced general bounds for subspace codes \cite{ShorLaflamme1997,RainsWeight1998,Rains1999,AshikhminLitsyn1999}. Li and Xing proved the Hamming bound at every fixed distance for sufficiently large block length \cite{LiXing2009}; later work covered broad CSS and $q$-ary regimes \cite{SarvepalliKlappenecker2010}, excluded violations for $d<127$ \cite{Dallas2022}, and established stronger Hamming-like inequalities for classes of degenerate stabilizer codes \cite{NemecTansuwannont2023}. Any counterexample was therefore forced into an increasingly exceptional finite-length region. But the conjecture is pointwise in $(n,d)$: a large-block-length theorem or a result for a restricted algebraic family cannot exclude an isolated finite-length degenerate code. What remained was a uniform finite-length theorem valid for all block lengths and all binary quantum subspace codes, regardless of degeneracy. We prove it here.

\begin{theorem}\label{thm:main}
Every exact binary quantum subspace code $((n,K,d))$ with $K>1$ satisfies Eq.~\eqref{eq:qhb}.
\end{theorem}

The restriction $K>1$ is necessary: a one-dimensional code carries no unknown logical state and can be recovered by simply re-preparing that state. For every nontrivial code, the theorem shows that exact error equivalences may change how error sectors overlap, but not the largest dimension that can be protected at fixed $n$ and $t$. It is therefore a genuinely finite-length converse for the binary problem. The conclusion applies directly to nonadditive codes, rather than following from stabilizer structure, and the bound is attained by quantum Hamming codes at the known perfect parameters.

Rather than trying to restore a disjoint-sphere picture, the proof measures the entire overlap pattern exactly. The key quantity compares two descriptions of the same collision geometry: a Hamming-ball intersection number in Pauli error space and a squared Lloyd polynomial in the Fourier-dual association scheme. The Li--Xing linear program turns the Hamming bound into the statement that this normalized collision ratio never exceeds one. Monotonicity in block length places the worst ratio at the shortest length allowed by the Rains shadow bound; monotonicity in center separation then reduces the boundary problem to its nearest configurations. The last estimate is local: a node--edge decomposition organizes all collision types along a one-dimensional chain, and a charging inequality controls every edge contribution. Thus a global coding-theoretic question is reduced to uniform local combinatorial inequalities.

\textit{Fourier reduction.---} It suffices to prove the case $d=2t+1$, since a code of distance $2t+2$ also has distance at least $2t+1$. The proof uses the exact correspondence between Pauli errors modulo phase and the additive Hamming space $\Ffour^n$. This retains all degeneracy-induced collisions while replacing operator algebra by finite geometry. We use the standard Krawtchouk normalization for the quaternary Hamming association scheme \cite{Delsarte1973,MacWilliamsSloane1977}. For $\wt z=s$, define
\begin{equation}
\begin{aligned}
 P_i(s;n)&=\sum_{j=0}^{i}(-1)^j3^{i-j}
 \binom{s}{j}\binom{n-s}{i-j},\\
 \Lloyd_t^{(n)}(s)&=\sum_{i=0}^{t}P_i(s;n),\\
 c_t(n,s)&=\bigl|\Ball_t^{(n)}\cap
 (\Ball_t^{(n)}+z)\bigr|,
\end{aligned}
\label{eq:definitions}
\end{equation}
where $\Ball_t^{(n)}=\{x:\wt x\le t\}$. The character sum over the weight-$i$ shell is $P_i(s;n)$. Hence the Fourier transform of $\one_{\Ball_t^{(n)}}$ at any character of weight $i$ is $\Lloyd_t^{(n)}(i)$. Its autocorrelation at $z$ is $c_t(n,s)$, and unnormalized Fourier inversion gives
\begin{equation}
 \sum_{i=0}^{n}\Lloyd_t^{(n)}(i)^2P_i(s;n)
 =4^n c_t(n,s).
 \label{eq:fourier}
\end{equation}
The factor $4^n$ comes from unnormalized Fourier inversion, and the square of the Lloyd polynomial comes from the autocorrelation identity. Full conventions and the normalization check are given in Supplemental Material (SM), Sec.~S1.

Set $f_i=\Lloyd_t^{(n)}(i)^2$ and $f(x)=\sum_i f_iP_i(x;n)$. With $S=\{0,\ldots,2t\}$, Eq.~\eqref{eq:fourier} gives $f(s)=4^nc_t(n,s)$. The two balls are disjoint for $s>2t$, so $f(s)=0$ outside $S$, whereas $f(0)/f_0=4^n/\Vol_t(n)$. The coefficients $f_i$ are nonnegative, and they are strictly positive for indices in $S$ by the first-zero estimate in SM Sec.~S3. The Li--Xing linear-programming theorem \cite{LiXing2009} therefore reduces Eq.~\eqref{eq:qhb} to a single family of normalized intersection inequalities,
\begin{equation}
 R_{n,t}(s):=
 \frac{\Vol_t(n)c_t(n,s)}{\Lloyd_t^{(n)}(s)^2}
 \le1,
 \qquad 1\le s\le2t.
 \label{eq:Rtarget}
\end{equation}
The numerator is the collision multiplicity of two radius-$t$ error balls at separation $s$, while the denominator is the corresponding squared Fourier response. Thus $R_{n,t}(s)\le1$ asserts that no overlap is large enough, relative to its spectral weight, to move the linear-programming maximum away from the origin. The remaining Li--Xing hypotheses follow from this same geometry. Since $d\ge2t+1$, the set $S$ lies in $\{0,\ldots,d-1\}$. The coefficients are squares and hence nonnegative; strict positivity on $S$ follows from $\Lloyd_t^{(n)}(0)=\Vol_t(n)>0$ and the first-zero estimate for $P_t(s-1;n-1)$. Outside $S$, the balls are disjoint, so the required sign condition holds with equality. Condition~\eqref{eq:Rtarget} then places the maximum Li--Xing ratio at $s=0$, and the prefactor $2^{-n}$ converts $4^n/\Vol_t(n)$ into $2^n/\Vol_t(n)$.

\textit{Two monotonicity reductions.---} The ratio $R_{n,t}(s)$ still depends on both block length and center separation. Estimating the full two-parameter range directly would obscure the structure, so we remove the parameters one at a time. The first reduction identifies the most difficult block length. The ball volume, ball intersection, and Lloyd polynomial obey parallel length recursions: in each case, appending a coordinate outside the support of the center difference contributes the same-radius term plus three times the radius-$(t-1)$ term. Their relative increments satisfy the following sandwich.

\begin{lemma}[Ratio sandwich]\label{lem:sandwich}
If $m\ge6t-2$ and $1\le s\le2t$, then
\begin{equation}
 \frac{c_{t-1}(m,s)}{c_t(m,s)}
 \le\frac{\Vol_{t-1}(m)}{\Vol_t(m)}
 \le\frac{\Lloyd_{t-1}^{(m)}(s)}
          {\Lloyd_t^{(m)}(s)}.
 \label{eq:sandwich}
\end{equation}
\end{lemma}

For the first inequality, fix the local pattern on the $s$ differing coordinates. Every type contributes a positive multiple of $\Vol_{t-h}(m-s)$, and its normalization by $\Vol_t(m)$ increases with the radius. For the second, the shift identity $\Lloyd_t^{(m)}(s)=P_t(s-1;m-1)$ and the smallest-zero bound imply that $P_{t-1}/P_t$ is increasing on $0\le s-1\le2t-1$. Quantitatively,
\[
 \frac{\Vol_{t-1}(m)}{\Vol_t(m)}
 \le\frac{t}{3(m-t+1)}
 <\frac{t}{3(m-t)}
 \le\frac{\Lloyd_{t-1}^{(m)}(s)}
          {\Lloyd_t^{(m)}(s)}.
\]
The two inequalities have different origins. The first is type-by-type. If the behavior on the $s$ differing coordinates consumes maximal local distance $h$, the remaining contribution is $\Vol_{t-h}(m-s)$. The adjacent-shell ratios in the smaller space are bounded by the corresponding ratios in the full space when $m\ge2t-1$; after summation this gives monotonicity of $\Vol_{t-h}(m-s)/\Vol_t(m)$ in $t$. For the second inequality, orthonormalizing the Krawtchouk recurrence produces a $t\times t$ Jacobi matrix. Weyl's inequality bounds its least eigenvalue, and therefore the smallest zero $\xi_1$ of $P_t(x;N)$, by
\[
 \xi_1\ge\frac{3N-2(t-1)-2\sqrt{3(t-1)(N-t+2)}}4>2t-1
\]
when $N\ge6t-3$. The partial-fraction expansion of $P_{t-1}/P_t$ has positive residues, so this ratio is increasing to the left of $\xi_1$. These arguments, including all endpoint cases, are in SM Secs.~S2--S3.

Let $x,y,z$ denote the three ratios in Eq.~\eqref{eq:sandwich} at length $n-1$. The parallel recursions give
\begin{equation}
 \frac{R_{n,t}(s)}{R_{n-1,t}(s)}
 =\frac{(1+3x)(1+3y)}{(1+3z)^2}\le1.
 \label{eq:lengthmono}
\end{equation}
The Rains shadow bound gives $n\ge6t-1$ for every nontrivial binary code correcting $t$ errors \cite{Rains1999}. Iterating Eq.~\eqref{eq:lengthmono} therefore reduces the problem to $n_0=6t-1$. Conceptually, increasing the ambient block length cannot create a new violation: the collision multiplicity grows no faster than the spectral denominator that controls it. Any counterexample would already have to occur at the shortest admissible length.

At the boundary length $n_0=6t-1$, we next determine which relative position of the two error balls is most dangerous. Once $s=1$ is treated directly, we show that the normalized ratio decreases as the centers move apart. Write $C_s=c_t(n_0,s)$ and $L_s=\Lloyd_t^{(n_0)}(s)$. Let $I_s$ be the intersection of the two exact-weight-$t$ shells in length $n_0-1$, and put $M_s=\Lloyd_{t-1}^{(n_0-1)}(s)$. The shell intersection is nonempty for $2\le s\le2t-1$: on the support of the center difference, split the coordinates as evenly as possible between the two centers and use the remaining weight outside the support. A one-coordinate classification and the Krawtchouk first-difference identity give
\begin{equation}
 C_{s+1}=C_s-I_s,
 \qquad
 L_{s+1}=L_s-4M_s.
 \label{eq:sdiff}
\end{equation}
The first equality says that the points lost when the centers move one unit farther apart are precisely the common outer-boundary points. The second follows from $P_i(s+1;n)-P_i(s;n)=-4P_{i-1}(s;n-1)$.

For $u_s=M_s/L_s$, the positivity of the Lloyd values yields
\begin{equation}
\begin{aligned}
 R_{n_0,t}(s+1)\le R_{n_0,t}(s)
 &\Longleftrightarrow
 \frac{I_s}{C_s}\ge8u_s-16u_s^2,\\
 0<u_s&\le\frac{t}{18t-5s-1}.
\end{aligned}
\label{eq:decrement}
\end{equation}
The estimate for the decrement comes from a degree recurrence. With $N=6t-2$, $x=s-1$, $L=P_t(x;N)$, and $M=P_{t-1}(x;N-1)$, differentiation of the generating function and the first-difference identity give $tL=(3N-4x)M-4xQ$, where $Q=P_{t-2}(x-1;N-2)$. To control $Q$, set $D=N-2$ and $y=x-1$. The degree recurrence implies
\[
\begin{aligned}
 \frac{P_j(y;D)}{P_{j-1}(y;D)}
 &=\frac{3D-2j+2-4y}{j}\\
 &\quad-\frac{3(D-j+2)}{j}
 \frac{P_{j-2}(y;D)}{P_{j-1}(y;D)}.
\end{aligned}
\]
Starting from $P_0=1$ and $P_1=3D-4y$, induction gives $0<P_{j-1}/P_j\le1/5$ for $1\le j\le t-1$. A dimension-shift identity then yields $0<Q\le M/4$, and hence $tL\ge(18t-5s-1)M$; see SM Sec.~S4. Because $8u-16u^2$ is increasing on $[0,1/4]$, Eq.~\eqref{eq:decrement} follows from the purely combinatorial estimate
\begin{equation}
 \frac{C_s}{I_s}\le T_{t,s}:=
 \frac{(18t-5s-1)^2}
 {8t(16t-5s-1)}.
 \label{eq:Ttarget}
\end{equation}

\textit{Node--edge charging.---} We are left with the combinatorial estimate in Eq.~\eqref{eq:Ttarget}. The difficulty is that the full intersection $C_s$ contains points with many local imbalances, whereas $I_s$ records only the simultaneous outer boundary. The following decomposition compares them without discarding any collision type. Put $r=2t-s$ and $N=6t-2-s$. On the $s$ coordinates where the centers differ, classify a vector according to whether it equals the first center, the second center, or one of the two symbols different from both. Let $c$ count the third type and let $\delta$ be the imbalance between the first two types. The larger local distance is $(s+c+\delta)/2$, leaving outside radius $(r-c-\delta)/2$. Even $\delta$ produces nodes; odd $\delta$ produces edges between neighboring nodes. The parity split is what turns a high-dimensional intersection problem into a one-dimensional chain.

For $c\equiv s\pmod{2}$, the double-boundary node weight is
\[
 D_c=\binom{s}{c}2^c\binom{s-c}{(s-c)/2}
 3^{(r-c)/2}\binom{N}{(r-c)/2}>0.
\]
Writing $\mathcal C$ and $\mathcal E$ for the admissible node and edge sets, the exact classification is
\begin{equation}
 I_s=\sum_{c\in\mathcal C}D_c,
 \qquad
 C_s=\sum_{c\in\mathcal C}\alpha_cD_c
      +\sum_{e\in\mathcal E}\beta_eB_e.
 \label{eq:chain}
\end{equation}
For a fixed $c$, the allowed imbalance has the same parity as $s-c$. Writing an even imbalance as $2j$ produces the node coefficient $\alpha_c$; writing an odd imbalance as $2j+1$ produces the edge coefficient $\beta_e$. The binomial multiplicity records the two possible orientations of a nonzero imbalance, while the remaining coordinates contribute a Hamming-ball volume of radius reduced by $j$. Thus $\alpha_c$ sums all even imbalances based at node $c$, and $\beta_eB_e$ sums all odd imbalances based at edge $e$. Their finite formulas are in SM Sec.~S5, Eqs.~(S41) and (S43), and Eqs.~(S38)--(S44) verify that no type is omitted or counted twice. For an ordinary edge $e\ge1$, let $\rho_e=(r-e-1)/2$. Direct cancellation gives
\begin{equation}
\begin{aligned}
 B_e^2&=\gamma_eD_{e-1}D_{e+1},\\
 \gamma_e&=\frac{(e+1)(\rho_e+1)}{3e(N-\rho_e)},\qquad e\ge1,\\
 B_0&=\frac{D_1}{s+1}.
\end{aligned}
\label{eq:edge}
\end{equation}
The factors in $\gamma_e$ have a direct combinatorial origin: $(e+1)/e$ comes from choosing the third-type coordinates, $(\rho_e+1)/[3(N-\rho_e)]$ is the adjacent-shell ratio, and all central-binomial and powers-of-two factors cancel. The case $e=0$ is the unique one-sided edge and occurs only for odd $s$. Applying $\sqrt{D_{e-1}D_{e+1}}\le(D_{e-1}+D_{e+1})/2$ charges half of each ordinary edge coefficient to each endpoint. Summing these charged inequalities gives $C_s\le\sum_c\Lambda_cD_c$. Consequently, if every local load $\Lambda_c$ is at most $T_{t,s}$, then $C_s\le T_{t,s}I_s$, which is Eq.~\eqref{eq:Ttarget}.

A geometric shell-tail estimate (SM Sec.~S6) gives the uniform bounds
\begin{equation}
\begin{aligned}
 \alpha_c&\le A:=
 \frac{3(r+4t-1)(2r+12t-3)}{(r+12t-3)^2},\\
 \beta_e&\le B:=
 \frac{9(r+8t)(r+4t-1)}{(r+12t+1)^2},\\
 \gamma_e&\le G:=
 \frac{2r}{3(r+8t-2)}.
\end{aligned}
\label{eq:ABG}
\end{equation}
Every node away from the special endpoint therefore has load at most $A+B\sqrt{G}$. If $s$ is odd, the left endpoint also receives the special edge $e=0$; its extra coefficient is at most $H/(s+1)$, where
\[
 H=\frac{9(r+4t-1)(r+8t-1)}{(r+12t-1)^2},
\]
and it receives only half of one ordinary-edge charge. Exact positive-coefficient certificates in SM Sec.~S7 prove
\begin{equation}
 A+B\sqrt{G}\le T_{t,s},
 \qquad
 A+\frac{H}{s+1}+\frac{1}{2}B\sqrt{G}\le T_{t,s}.
 \label{eq:certificates}
\end{equation}
For the internal-node inequality, the substitutions $x=r-2$, $y=s-4$ turn both $T-A$ and $(T-A)^2-B^2G$ into rational functions with positive denominators and strictly positive-coefficient numerators. For the odd endpoint, $x=r-3$, $y=s-7$ does the same for $E=T-A-H/(s+1)$ and $E^2-B^2G/4$. The recorded coefficient tables contain $105$, $35$, and $207$ positive terms for the three nontrivial certificate polynomials and are reconstructed by the accompanying exact-arithmetic script.

The uniform argument covers even $s\ge4$ with $r\ge2$ and odd $s\ge7$ with $r\ge3$. The remaining cases are short enough to classify exactly. For example, with $w_j(N):=3^j\binom{N}{j}$,
\[
\begin{aligned}
 C_2&=2\Vol_{t-1}(6t-3)+14\Vol_{t-2}(6t-3),\\
 I_2&=2w_{t-1}(6t-4)+4w_{t-2}(6t-4),
\end{aligned}
\]
with analogous two- and three-shell formulas for $s=3$ and $s=5$. When $r=1$, the chain contains only node $c=1$ and the special edge, giving $C_{2t-1}/I_{2t-1}=(t+1)/t$. Shell-tail bounds verify Eq.~\eqref{eq:Ttarget} for all of these families (SM Sec.~S8). Parity of $r=2t-s$ shows that the six regions are disjoint and exhaust $2\le s\le2t-1$. The same section proves the starting estimates, so that
\begin{equation}
\begin{aligned}
 R_{6t-1,t}(1)&<\frac{1}{2},\qquad
 R_{6t-1,t}(2)<1,\\
 R_{6t-1,t}(2)&\ge R_{6t-1,t}(3)
 \ge\cdots\ge R_{6t-1,t}(2t).
\end{aligned}
\label{eq:boundary}
\end{equation}
For $t=1$, the two values are $4/9$ and $1/2$. Hence Eq.~\eqref{eq:Rtarget} holds at the Rains boundary, and Eq.~\eqref{eq:lengthmono} extends it to every admissible length. The Li--Xing reduction proves Theorem~\ref{thm:main}.

\textit{Implications and outlook.---} Theorem~\ref{thm:main} settles the quantum Hamming-bound problem for exact binary subspace codes. Degeneracy can identify distinct correctable errors and invalidate the standard sphere-packing proof, but it cannot produce a code with $K\Vol_t(n)>2^n$. Because the argument applies directly to arbitrary quantum subspaces, the conclusion includes nonadditive codes and does not rely on stabilizer structure. The known perfect quantum Hamming codes show, moreover, that this universal finite-length bound is tight.

The theorem does not make degeneracy unimportant. Degenerate structure can change the syndrome map, reorganize low-weight errors, and be valuable in code design. What it excludes is one specific advantage: degeneracy under exact correction cannot be converted into additional logical dimension at fixed block length and correction radius. The overlap of physical error sectors is real, but its apparent saving is offset by constraints in the dual weight-enumerator geometry.

The proof makes this compensation explicit. Degeneracy is retained exactly in the ball-intersection numbers $c_t(n,s)$ rather than removed by a reduction to nondegenerate or stabilizer codes. Fourier duality compares these collision multiplicities with squared Lloyd coefficients. Length monotonicity places the worst normalized collision at the shortest admissible block length, while separation monotonicity shows that moving the centers farther apart cannot increase the critical ratio. The node--edge chain supplies the local mechanism behind the final step. Thus the failure of elementary dimension counting is repaired not by finding a new set of disjoint error sectors, but by proving that the complete overlap pattern obeys a stronger global balance law.

The binary and exact settings enter essentially. The factors $3$ and $4$ arise from the Pauli alphabet, and the boundary $n=6t-1$ is specific to binary subspace codes. Approximate correction, subsystem codes, and entanglement-assisted codes obey different operational constraints and require different counting objects. A $q$-ary extension would preserve the Fourier viewpoint, but the shell ratios, zero-location estimate, boundary length, and local certificates would all have to be rebuilt.

Computer algebra is used only to expand four explicit rational identities and verify positivity of their integer coefficient tables. No numerical parameter scan, semidefinite program, or numerical optimization enters the proof. The accompanying exact-arithmetic source reconstructs these certificates and checks the exceptional families, making the endpoint algebra independently auditable.

Two features of the argument may be reusable. First, a quantum linear-programming polynomial is realized as an exact intersection problem in a classical association scheme. Second, the resulting global inequality is converted into local loads on a chain. Similar Fourier--intersection correspondences may help with other finite-length quantum coding bounds for which degeneracy or nonadditivity obstructs a direct packing argument. The positive-coefficient certificates also suggest an underlying probabilistic or representation-theoretic explanation that remains to be found.

\begin{acknowledgments}
This work is supported by the Quantum Science and Technology-National Science and Technology Major Project (Grant No. 2024ZD0301000), and the National Natural Science Foundation of China (Grant No. 12275136).
\end{acknowledgments}


\begin{thebibliography}{99}
\bibitem{Shor1995} P.~W. Shor, ``Scheme for reducing decoherence in quantum computer memory,'' \doilink{10.1103/PhysRevA.52.R2493}{Phys. Rev. A \textbf{52}, R2493 (1995)}.
\bibitem{CalderbankShor1996} A.~R. Calderbank and P.~W. Shor, ``Good quantum error-correcting codes exist,'' \doilink{10.1103/PhysRevA.54.1098}{Phys. Rev. A \textbf{54}, 1098 (1996)}.
\bibitem{Steane1996} A.~M. Steane, ``Error correcting codes in quantum theory,'' \doilink{10.1103/PhysRevLett.77.793}{Phys. Rev. Lett. \textbf{77}, 793 (1996)}.
\bibitem{EkertMacchiavello1996} A.~Ekert and C.~Macchiavello, ``Quantum error correction for communication,'' \doilink{10.1103/PhysRevLett.77.2585}{Phys. Rev. Lett. \textbf{77}, 2585 (1996)}.
\bibitem{Laflamme1996} R.~Laflamme, C.~Miquel, J.~P. Paz, and W.~H. Zurek, ``Perfect quantum error correcting code,'' \doilink{10.1103/PhysRevLett.77.198}{Phys. Rev. Lett. \textbf{77}, 198 (1996)}.
\bibitem{KnillLaflamme1997} E.~Knill and R.~Laflamme, ``Theory of quantum error-correcting codes,'' \doilink{10.1103/PhysRevA.55.900}{Phys. Rev. A \textbf{55}, 900 (1997)}.
\bibitem{Gottesman1996} D.~Gottesman, ``Class of quantum error-correcting codes saturating the quantum Hamming bound,'' \doilink{10.1103/PhysRevA.54.1862}{Phys. Rev. A \textbf{54}, 1862 (1996)}.
\bibitem{Aly2007} S.~A. Aly, ``A note on quantum Hamming bound,'' \doilink{10.48550/arXiv.0711.4603}{arXiv:0711.4603 [quant-ph] (2007)}.
\bibitem{LiXing2009} Z.~Li and L.~Xing, ``Progress on problem about quantum Hamming bound for impure quantum codes,'' \doilink{10.48550/arXiv.0907.3802}{arXiv:0907.3802 [quant-ph] (2009)}.
\bibitem{SarvepalliKlappenecker2010} P.~K. Sarvepalli and A.~Klappenecker, ``Degenerate quantum codes and the quantum Hamming bound,'' \doilink{10.1103/PhysRevA.81.032318}{Phys. Rev. A \textbf{81}, 032318 (2010)}.
\bibitem{Dallas2022} E.~Dallas, F.~Andreadakis, and D.~A. Lidar, ``No $((n,K,d<127))$ code can violate the quantum Hamming bound,'' \doilink{10.1109/MBITS.2023.3262219}{IEEE BITS Inf. Theory Mag. \textbf{2}, 33 (2022)}.
\bibitem{CalderbankRainsShorSloane1998} A.~R. Calderbank, E.~M. Rains, P.~W. Shor, and N.~J.~A. Sloane, ``Quantum error correction via codes over GF(4),'' \doilink{10.1109/18.681315}{IEEE Trans. Inf. Theory \textbf{44}, 1369 (1998)}.
\bibitem{RainsHardinShorSloane1997} E.~M. Rains, R.~H. Hardin, P.~W. Shor, and N.~J.~A. Sloane, ``A nonadditive quantum code,'' \doilink{10.1103/PhysRevLett.79.953}{Phys. Rev. Lett. \textbf{79}, 953 (1997)}.
\bibitem{SmolinSmithWehner2007} J.~A. Smolin, G.~Smith, and S.~Wehner, ``Simple family of nonadditive quantum codes,'' \doilink{10.1103/PhysRevLett.99.130505}{Phys. Rev. Lett. \textbf{99}, 130505 (2007)}.
\bibitem{ShorLaflamme1997} P.~W. Shor and R.~Laflamme, ``Quantum analog of the MacWilliams identities in classical coding theory,'' \doilink{10.1103/PhysRevLett.78.1600}{Phys. Rev. Lett. \textbf{78}, 1600 (1997)}.
\bibitem{RainsWeight1998} E.~M. Rains, ``Quantum weight enumerators,'' \doilink{10.1109/18.681316}{IEEE Trans. Inf. Theory \textbf{44}, 1388 (1998)}.
\bibitem{Rains1999} E.~M. Rains, ``Quantum shadow enumerators,'' \doilink{10.1109/18.796376}{IEEE Trans. Inf. Theory \textbf{45}, 2361 (1999)}.
\bibitem{AshikhminLitsyn1999} A.~Ashikhmin and S.~Litsyn, ``Upper bounds on the size of quantum codes,'' \doilink{10.1109/18.761270}{IEEE Trans. Inf. Theory \textbf{45}, 1206 (1999)}.
\bibitem{NemecTansuwannont2023} A.~Nemec and T.~Tansuwannont, ``A Hamming-like bound for degenerate stabilizer codes,'' \doilink{10.48550/arXiv.2306.00048}{arXiv:2306.00048 [quant-ph] (2023)}.
\bibitem{Delsarte1973} P.~Delsarte, ``An algebraic approach to the association schemes of coding theory,'' Philips Res. Rep. Suppl. \textbf{10} (1973).
\bibitem{MacWilliamsSloane1977} F.~J. MacWilliams and N.~J.~A. Sloane, \textit{The Theory of Error-Correcting Codes} (North-Holland, Amsterdam, 1977).
\end{thebibliography}
\end{document}


\title{Supplemental Material for ``Degeneracy Cannot Violate the Quantum Hamming Bound''}

\author{Yu-Xuan Zhang}
\affiliation{School of Physics, Nankai University, Tianjin 300071, People's Republic of China}

\author{Jing-Ling Chen}
\email{chenjl@nankai.edu.cn}
\affiliation{Theoretical Physics Division, Chern Institute of Mathematics, Nankai University, Tianjin 300071, People's Republic of China}

\date{June 12, 2026}
\maketitle

This Supplemental Material records normalization checks, the complete
combinatorial decomposition, the endpoint calculations, and the exact
algebraic certificates used in the main text. Equation numbers prefixed by
S are local to this document.

\smsection{The Li--Xing reduction and Fourier normalization}{sec:S1}

We work in the additive group $\Ffour^n$, identified with the $n$-qubit
Pauli group modulo phases, and write $\wt(\,\cdot\,)$ for the Hamming
weight. Define
\begin{equation*}
 \Ball_t^{(n)}
 =
 \{x\in\Ffour^n:\wt(x)\le t\},
 \qquad
 \Vol_t(n)
 =
 |\Ball_t^{(n)}|
 =
 \sum_{j=0}^{t}3^j\binom{n}{j}.
\end{equation*}
For a vector $z\in\Ffour^n$ of weight $s$, let
\begin{equation*}
 c_t(n,s)
 =
 \left|
 \Ball_t^{(n)}
 \cap
 \bigl(\Ball_t^{(n)}+z\bigr)
 \right|.
\end{equation*}
This quantity depends only on $s$.

The quaternary Krawtchouk polynomial is
\begin{equation}
 P_i(s;n)=\sum_{j=0}^{i}(-1)^j3^{i-j}
 \binom{s}{j}\binom{n-s}{i-j}.
 \tag{S1}
\end{equation}
Let $\overline{a}=a^2$ denote conjugation in $\Ffour$, and define the
trace-symplectic pairing and the associated additive character by
\begin{equation*}
 \langle u,z\rangle_{\mathrm{s}}
 =
 \operatorname{Tr}_{\Ffour/\mathbb{F}_2}
 \left(\sum_{k=1}^{n}u_k\overline{z_k}\right),
 \qquad
 \chi_u(z)
 =
 (-1)^{\langle u,z\rangle_{\mathrm{s}}}.
\end{equation*}
The character sum over a weight-$i$ shell is
\begin{equation}
 P_i(\wt z;n)=\sum_{\substack{u\in\Ffour^n\\ \wt u=i}}\chi_u(z).
 \tag{S2}
\end{equation}
Indeed, a nonzero coordinate of $u$ outside the support of $z$ contributes
a factor $3$ after summation over its three possible values, whereas a
nonzero coordinate of $u$ on the support of $z$ contributes a factor $-1$.

Define the Lloyd polynomial by
\begin{equation*}
 \Lloyd_t^{(n)}(s)
 =
 \sum_{j=0}^{t}P_j(s;n),
\end{equation*}
and let
\begin{equation*}
 g(x)=\one_{\Ball_t^{(n)}}(x),
 \qquad
 \widehat g(u)
 =
 \sum_{x\in\Ffour^n}g(x)\chi_u(x).
\end{equation*}
If $\wt u=i$, then Eq.~(S2) gives
\begin{equation}
 \widehat g(u)
 =
 \sum_{j=0}^{t}P_j(i;n)
 =
 \Lloyd_t^{(n)}(i).
 \tag{S3}
\end{equation}

For the unnormalized Fourier transform
\begin{equation*}
 \widehat h(u)
 =
 \sum_{x\in\Ffour^n}h(x)\chi_u(x),
\end{equation*}
the inversion formula is
\begin{equation*}
 h(z)
 =
 4^{-n}
 \sum_{u\in\Ffour^n}\widehat h(u)\chi_u(z).
\end{equation*}
Define the autocorrelation of $g$ by
\begin{equation*}
 A_g(z)
 =
 \sum_{x\in\Ffour^n}g(x)g(x+z).
\end{equation*}
If $\wt z=s$, then
\begin{equation*}
 A_g(z)
 =
 \left|
 \Ball_t^{(n)}
 \cap
 \bigl(\Ball_t^{(n)}+z\bigr)
 \right|
 =
 c_t(n,s).
\end{equation*}
Moreover,
\begin{align*}
 \widehat{A_g}(u)
 &=
 \sum_{z\in\Ffour^n}
 \sum_{x\in\Ffour^n}
 g(x)g(x+z)\chi_u(z)\\
 &=
 \widehat g(u)\overline{\widehat g(u)}
 =
 |\widehat g(u)|^2.
\end{align*}
Fourier inversion, followed by Eqs.~(S2) and (S3), therefore gives
\begin{align}
 4^nc_t(n,s)
 &=
 \sum_{u\in\Ffour^n}
 |\widehat g(u)|^2\chi_u(z)
 \notag\\
 &=
 \sum_{i=0}^{n}
 \Lloyd_t^{(n)}(i)^2
 \sum_{\substack{u\in\Ffour^n\\ \wt u=i}}
 \chi_u(z)
 \notag\\
 &=
 \sum_{i=0}^{n}
 \Lloyd_t^{(n)}(i)^2P_i(s;n).
 \tag{S4}
\end{align}
This proves the Fourier--intersection identity, including the normalization
factor $4^n$.

For completeness, we recall the binary specialization of the Li--Xing
linear-programming theorem \cite{LiXing2009}. Let
\begin{equation*}
 T=\{0,\ldots,n\},
 \qquad
 f(x)=\sum_{i=0}^{n}f_iP_i(x;n).
\end{equation*}
Suppose that $S$ is a nonempty subset of $\{0,\ldots,d-1\}$ and that
\begin{equation*}
 f_i>0
 \quad(i\in S),
 \qquad
 f_i\ge0
 \quad(i\in T\setminus S),
\end{equation*}
while
\begin{equation*}
 f(x)\le0
 \quad(x\in T\setminus S).
\end{equation*}
Then every binary $((n,K,d))$ quantum code satisfies
\begin{equation}
 K\le2^{-n}\max_{x\in S}\frac{f(x)}{f_x}.
 \tag{S5}
\end{equation}

We now apply this theorem. The case $t=0$ is just the trivial dimension
bound, so assume $t\ge1$, $d\ge2t+1$, and $n\ge6t-1$. Set
\begin{equation*}
 S=\{0,\ldots,2t\}.
\end{equation*}
Then $S\subseteq\{0,\ldots,d-1\}$ and $S\subseteq T$. Choose
\begin{equation}
 f_i=\Lloyd_t^{(n)}(i)^2.
 \tag{S6}
\end{equation}
Clearly $f_i\ge0$ for every $i\in T$. We next verify the strict positivity
required on $S$. At $i=0$,
\begin{equation*}
 \Lloyd_t^{(n)}(0)
 =
 \Vol_t(n)>0.
\end{equation*}
For $1\le i\le2t$, the shift identity in Eq.~(S19) gives
\begin{equation*}
 \Lloyd_t^{(n)}(i)
 =
 P_t(i-1;n-1).
\end{equation*}
Since $n\ge6t-1$, the first-zero estimate in
Lemma~\ref{lem:Szero} places the smallest zero of
$P_t(x;n-1)$ strictly to the right of $2t-1$. Because
\begin{equation*}
 P_t(0;n-1)
 =
 3^t\binom{n-1}{t}>0,
\end{equation*}
it follows that
\begin{equation*}
 \Lloyd_t^{(n)}(i)>0,
 \qquad
 1\le i\le2t.
\end{equation*}
Thus $f_i>0$ for every $i\in S$.

By Eq.~(S4),
\begin{equation*}
 f(s)=4^nc_t(n,s).
\end{equation*}
If $s>2t$, two radius-$t$ Hamming balls whose centers are separated by
distance $s$ are disjoint. Hence
\begin{equation*}
 f(s)=0
 \qquad
 (s\in T\setminus S),
\end{equation*}
so the sign condition in the Li--Xing theorem holds with equality.

At $s=0$, the two balls coincide, and therefore
\begin{equation}
 f(0)=4^n\Vol_t(n),
 \qquad
 f_0=\Vol_t(n)^2.
 \tag{S7}
\end{equation}
Consequently,
\begin{equation*}
 \frac{f(0)}{f_0}
 =
 \frac{4^n}{\Vol_t(n)}.
\end{equation*}
For $1\le s\le2t$, the condition that the value at $s=0$ dominate the
corresponding ratio in Eq.~(S5) is
\begin{equation*}
 \frac{4^nc_t(n,s)}
 {\Lloyd_t^{(n)}(s)^2}
 \le
 \frac{4^n}{\Vol_t(n)}.
\end{equation*}
Equivalently,
\begin{equation}
 R_{n,t}(s):=
 \frac{\Vol_t(n)c_t(n,s)}
 {\Lloyd_t^{(n)}(s)^2}
 \le1.
 \tag{S8}
\end{equation}
Therefore, if Eq.~(S8) holds for every $1\le s\le2t$, then
\begin{equation*}
 \max_{s\in S}\frac{f(s)}{f_s}
 =
 \frac{4^n}{\Vol_t(n)}.
\end{equation*}
Equation~(S5) then yields
\begin{equation*}
 K
 \le
 2^{-n}\frac{4^n}{\Vol_t(n)}
 =
 \frac{2^n}{\Vol_t(n)},
\end{equation*}
which is the binary quantum Hamming bound.

\smsection{Shell-tail inequalities and the overlap-radius comparison}{sec:S2}

Write
\begin{equation}
 w_j(N)=3^j\binom{N}{j},
 \qquad
 \Vol_j(N)=\sum_{\ell=0}^{j}w_\ell(N),
 \tag{S9}
\end{equation}
with $w_j=\Vol_j=0$ for $j<0$.

\begin{lemma}[Shell-tail bound]\label{lem:Sshell}
Let
\begin{equation*}
 q=\frac{j}{3(N-j+1)}<1.
\end{equation*}
For every $h\ge0$,
\begin{align}
 \frac{w_{j-h}(N)}{w_j(N)}&\le q^h,\tag{S10}\\
 \frac{\Vol_{j-1}(N)}{\Vol_j(N)}&\le q,\tag{S11}\\
 \frac{\Vol_{j-h}(N+1)}{w_j(N)}
 &\le q^h\frac{1+3q}{1-q}.
 \tag{S12}
\end{align}
\end{lemma}

\begin{proof}
The adjacent-shell ratio
\begin{equation}
 \frac{w_{\ell-1}(N)}{w_\ell(N)}
 =\frac{\ell}{3(N-\ell+1)}
 \tag{S13}
\end{equation}
increases with $\ell$.  Iteration gives Eq.~(S10).  Therefore
\begin{equation*}
 \Vol_{j-1}(N)\le w_j(N)\frac{q}{1-q},
\end{equation*}
which implies Eq.~(S11) after adding $w_j(N)$ to the denominator.  Finally,
\begin{equation*}
 \Vol_k(N+1)=\Vol_k(N)+3\Vol_{k-1}(N)
\end{equation*}
and Eq.~(S10) give Eq.~(S12) with $k=j-h$.
\end{proof}

We next compare the radius distribution of a ball intersection with the radius distribution of a single ball.

\begin{lemma}[Normalized intersections increase with radius]\label{lem:Soverlap}
If $m\ge2t-1$ and $1\le s\le\min(2t,m)$, then
\begin{equation}
 \frac{c_{t-1}(m,s)}{c_t(m,s)}
 \le\frac{\Vol_{t-1}(m)}{\Vol_t(m)}.
 \tag{S14}
\end{equation}
\end{lemma}

\begin{proof}
Fix a center difference $z$ of weight $s$.  On its support, let $a$ coordinates of $x$ equal $0$, let $b$ equal $z_i$, and let $c$ take one of the two remaining symbols.  Then $a+b+c=s$, the multiplicity is
\begin{equation*}
 \frac{s!}{a!b!c!}2^c,
\end{equation*}
and the larger of the two local distances is
\begin{equation}
 h=c+\max(a,b)=s-\min(a,b).
 \tag{S15}
\end{equation}
Consequently,
\begin{equation}
 c_t(m,s)=
 \sum_{a+b+c=s}\frac{s!}{a!b!c!}2^c
 \Vol_{t-h}(m-s).
 \tag{S16}
\end{equation}
It suffices to show that
\begin{equation*}
 Q_t(h):=\frac{\Vol_{t-h}(m-s)}{\Vol_t(m)}
\end{equation*}
increases with $t$ for every local type.  Set $r=t-h$ and $N=m-s$.  For $0\le p<r$,
\begin{equation}
 \frac{w_{r-p-1}(N)}{w_{r-p}(N)}
 \le
 \frac{w_{t-p-1}(m)}{w_{t-p}(m)}
 \tag{S17}
\end{equation}
is equivalent to
\begin{equation}
 h(m-t+p+1)\ge(t-p)(s-h).
 \tag{S18}
\end{equation}
Now $h\ge s/2$, hence $h\ge s-h$, while $m\ge2t-1$ gives $m-t+p+1\ge t-p$.  Thus Eq.~(S18) holds.  If $r<0$, the local type contributes zero to both sides; if $r=0$, then $Q_{t-1}(h)=0$ and the claim is immediate.  Assume $1\le r\le t$ and put
\[
 a_p=\frac{w_{r-p-1}(N)}{w_{r-p}(N)},\qquad
 b_p=\frac{w_{t-p-1}(m)}{w_{t-p}(m)}.
\]
Equation~(S17) says $a_p\le b_p$ for $0\le p<r$.  Hence
\begin{align*}
 \frac{\Vol_r(N)}{w_r(N)}
 &=1+a_0+a_0a_1+\cdots+a_0a_1\cdots a_{r-1}\\
 &\le1+b_0+b_0b_1+\cdots+b_0b_1\cdots b_{t-1}
 =\frac{\Vol_t(m)}{w_t(m)}.
\end{align*}
Consequently $w_r(N)/\Vol_r(N)\ge w_t(m)/\Vol_t(m)$, and therefore
\begin{align*}
 \frac{\Vol_r(N)}{\Vol_{r-1}(N)}
 &=\frac{1}{1-w_r(N)/\Vol_r(N)}\\
 &\ge\frac{1}{1-w_t(m)/\Vol_t(m)}
 =\frac{\Vol_t(m)}{\Vol_{t-1}(m)}.
\end{align*}
This is exactly $Q_t(h)\ge Q_{t-1}(h)$.  Summing Eq.~(S16) proves Eq.~(S14).
\end{proof}

\smsection{The Lloyd ratio and monotonicity in length}{sec:S3}

The Lloyd polynomial obeys the standard shift identity
\begin{equation}
 \Lloyd_t^{(n)}(s)=P_t(s-1;n-1),\qquad s\ge1.
 \tag{S19}
\end{equation}
We first locate the smallest zero of the polynomial on the right.

\begin{lemma}[First zero]\label{lem:Szero}
If $N\ge6t-3$, the smallest zero $\xi_1$ of $P_t(x;N)$ satisfies
\begin{equation}
 \xi_1>2t-1.
 \tag{S20}
\end{equation}
\end{lemma}

\begin{proof}
After orthonormalization and an alternating sign change, multiplication by $x$ on degrees $0,\ldots,t-1$ is represented by a symmetric Jacobi matrix whose diagonal and upper off-diagonal entries are
\begin{equation}
 b_i=\frac{3N-2i}{4},
 \qquad
 a_i=\frac{\sqrt{3(i+1)(N-i)}}4,
 \quad 0\le i\le t-2.
 \tag{S21}
\end{equation}
For $t=1$, the unique zero is $3N/4>1=2t-1$, so assume $t\ge2$.  The zeros of $P_t$ are the eigenvalues of this $t\times t$ principal block.  Write the block as $J=D+E$, with $D$ diagonal and $E$ containing the two off-diagonals.  Since $b_i$ decreases with $i$,
\[
 \lambda_{\min}(D)=b_{t-1}=\frac{3N-2(t-1)}4.
\]
Moreover,
\[
 (i+2)(N-i-1)-(i+1)(N-i)=N-2i-2>0
\]
for $0\le i\le t-3$ when $N\ge6t-3$.  Hence
$\max_i a_i=a_{t-2}$, and the maximum-row-sum bound gives
\[
 \|E\|\le2\max_i a_i
 =\frac{\sqrt{3(t-1)(N-t+2)}}2.
\]
Weyl's inequality therefore gives
\begin{equation}
 \xi_1\ge
 \frac{3N-2(t-1)-2\sqrt{3(t-1)(N-t+2)}}4.
 \tag{S22}
\end{equation}
The derivative of the right-hand side with respect to $N$ is
\[
 \frac{1}{4}\left(3-\sqrt{\frac{3(t-1)}{N-t+2}}\right)>0
\]
in the stated range.  It is therefore enough to set $N=6t-3$.  The assertion that the resulting bound exceeds $2t-1$ reduces to
\begin{equation*}
 8t-3>2\sqrt{3(t-1)(5t-1)},
\end{equation*}
whose squared difference is
\begin{equation*}
 (8t-3)^2-12(t-1)(5t-1)=4t^2+24t-3>0.
\end{equation*}
\end{proof}

To fix the normalization, define the monic polynomials
\[
 \pi_j(x)=(-1)^j\frac{j!}{4^j}P_j(x;N).
\]
For a monic orthogonal-polynomial sequence with simple zeros
$\xi_1<\cdots<\xi_t$, the consecutive-degree ratio has the exact expansion
\[
 \frac{\pi_{t-1}(x)}{\pi_t(x)}
 =\sum_{j=1}^{t}\frac{\omega_j}{x-\xi_j},
 \qquad \omega_j>0.
\]
Since $P_{t-1}/P_t=-(t/4)\,\pi_{t-1}/\pi_t$, this becomes
\begin{equation}
 \frac{P_{t-1}(x;N)}{P_t(x;N)}
 =\frac{t}{4}\sum_{j=1}^{t}\frac{\omega_j}{\xi_j-x}.
 \tag{S23}
\end{equation}
Consequently,
\[
 \frac{d}{dx}\frac{P_{t-1}(x;N)}{P_t(x;N)}
 =\frac{t}{4}\sum_{j=1}^{t}\frac{\omega_j}{(\xi_j-x)^2}>0
\]
for $x<\xi_1$.

\begin{lemma}[Volume-to-Lloyd comparison]\label{lem:Slloyd}
If $m\ge6t-2$ and $1\le s\le2t$, then
\begin{equation}
 \frac{\Vol_{t-1}(m)}{\Vol_t(m)}
 \le
 \frac{\Lloyd_{t-1}^{(m)}(s)}{\Lloyd_t^{(m)}(s)}.
 \tag{S24}
\end{equation}
\end{lemma}

\begin{proof}
Lemma~\ref{lem:Sshell} gives
\begin{equation}
 \frac{\Vol_{t-1}(m)}{\Vol_t(m)}
 \le\frac{t}{3(m-t+1)}.
 \tag{S25}
\end{equation}
Set $N=m-1$ and $x=s-1$.  Equations~(S19), (S20), and (S23) give
\begin{align}
 \frac{\Lloyd_{t-1}^{(m)}(s)}{\Lloyd_t^{(m)}(s)}
 &=\frac{P_{t-1}(x;N)}{P_t(x;N)}\\
 &\ge\frac{P_{t-1}(0;N)}{P_t(0;N)}
 =\frac{t}{3(m-t)}.
 \tag{S26}
\end{align}
Combining Eqs.~(S25) and (S26) proves the result.  The same argument also shows that all $\Lloyd_t^{(m)}(s)$ appearing here are positive.
\end{proof}

For $n\ge6t-1$ and $s\le2t<n$, choose a coordinate outside the support of the center difference.  The recursions
\begin{align}
 \Vol_t(n)&=\Vol_t(n-1)+3\Vol_{t-1}(n-1),\tag{S27}\\
 c_t(n,s)&=c_t(n-1,s)+3c_{t-1}(n-1,s),\tag{S28}\\
 \Lloyd_t^{(n)}(s)&=\Lloyd_t^{(n-1)}(s)
 +3\Lloyd_{t-1}^{(n-1)}(s)\tag{S29}
\end{align}
follow by classifying the last coordinate.  Put
\begin{equation*}
 x=\frac{\Vol_{t-1}}{\Vol_t},\qquad
 y=\frac{c_{t-1}}{c_t},\qquad
 z=\frac{\Lloyd_{t-1}}{\Lloyd_t},
\end{equation*}
with all terms evaluated at length $n-1$.  Lemmas~\ref{lem:Soverlap} and~\ref{lem:Slloyd} give $y\le x\le z$, and therefore
\begin{equation}
 \frac{R_{n,t}(s)}{R_{n-1,t}(s)}
 =\frac{(1+3x)(1+3y)}{(1+3z)^2}\le1.
 \tag{S30}
\end{equation}
Together with the Rains bound $n\ge6t-1$ for $K>1$ \cite{Rains1999}, this reduces the proof to $n=6t-1$.

\smsection{Separation differences and the Lloyd decrement}{sec:S4}

Fix $n_0=6t-1$ and write
\begin{equation*}
 C_s=c_t(n_0,s),\quad L_s=\Lloyd_t^{(n_0)}(s),\quad
 M_s=\Lloyd_{t-1}^{(n_0-1)}(s).
\end{equation*}
Let $A_{p,q}$ count vectors in the remaining $n_0-1$ coordinates whose distances from the two restricted centers are $p$ and $q$.  When the new coordinate is equal in the two centers,
\begin{equation*}
 C_s=\sum_{p\le t,q\le t}A_{p,q}
 +3\sum_{p\le t-1,q\le t-1}A_{p,q}.
\end{equation*}
When the new center symbols differ,
\begin{align*}
 C_{s+1}={}&\sum_{p\le t,q\le t-1}A_{p,q}
 +\sum_{p\le t-1,q\le t}A_{p,q}\nonumber\\
 &+2\sum_{p\le t-1,q\le t-1}A_{p,q}.
\end{align*}
Their difference is $A_{t,t}$, which is exactly
\begin{equation*}
 I_s=\left|\Shell_t^{(n_0-1)}\cap
 (\Shell_t^{(n_0-1)}+z)\right|.
\end{equation*}
Thus
\begin{equation}
 C_{s+1}=C_s-I_s.
 \tag{S31}
\end{equation}
The Krawtchouk difference identity
\begin{equation*}
 P_i(s+1;n)-P_i(s;n)=-4P_{i-1}(s;n-1)
\end{equation*}
gives
\begin{equation}
 L_{s+1}=L_s-4M_s.
 \tag{S32}
\end{equation}
Set $R_s=R_{n_0,t}(s)$ and $u_s=M_s/L_s$. Substituting Eqs.~(S31)--(S32) into the definition of $R$ yields
\begin{equation}
 R_{s+1}\le R_s
 \iff
 \frac{I_s}{C_s}\ge 8u_s-16u_s^2.
 \tag{S33}
\end{equation}

We now prove the bound on $u_s$.  Put
\begin{equation*}
 N=6t-2,\quad x=s-1,
 \quad L=P_t(x;N),
 \quad M=P_{t-1}(x;N-1).
\end{equation*}
Differentiating the generating function
\begin{equation*}
 (1+3z)^{N-x}(1-z)^x
\end{equation*}
and taking the coefficient of $z^{t-1}$ gives
\begin{equation*}
 tL=3(N-x)M-xP_{t-1}(x-1;N-1).
\end{equation*}
The difference identity gives
\begin{equation*}
 P_{t-1}(x-1;N-1)=M+4Q,
 \qquad Q=P_{t-2}(x-1;N-2),
\end{equation*}
so
\begin{equation}
 tL=(3N-4x)M-4xQ.
 \tag{S34}
\end{equation}

Set $D=N-2=6t-4$, $y=x-1=s-2$, and
\begin{equation*}
 q_j=\frac{P_{j-1}(y;D)}{P_j(y;D)}.
\end{equation*}
The degree recurrence is
\begin{equation*}
 jP_j(y;D)=(3D-2j+2-4y)P_{j-1}(y;D)
 -3(D-j+2)P_{j-2}(y;D),
\end{equation*}
whence
\begin{equation}
 \frac{1}{q_j}=
 \frac{3D-2j+2-4y}{j}
 -\frac{3(D-j+2)}{j}q_{j-1}.
 \tag{S35}
\end{equation}
The base case satisfies
\[
 0<q_1=\frac{1}{3D-4y}\le\frac{1}{10t}\le\frac{1}{5}.
\]
We prove simultaneously that $P_j(y;D)>0$ and
$0<q_j\le1/5$.  Suppose this holds through degree $j-1$.
The right-hand side of Eq.~(S35) is at least $5$ provided
\begin{equation*}
 12D-20y+4\ge32j.
\end{equation*}
For $y\le2t-3$ and $j\le t-1$, the left side is at least $32t+16$ and the right side at most $32t-32$.  The recurrence therefore gives
$P_j/P_{j-1}\ge5>0$, proving both $P_j>0$ and
$0<q_j\le1/5$ by induction.  With
\begin{equation*}
 A=P_{t-1}(y;D)>0,
\end{equation*}
we have $0<Q/A\le1/5$, while the dimension-shift recurrence gives
$M=A-Q\ge4A/5>0$.  Hence $Q\le M/4$.  Equation~(S34) now implies
\begin{equation*}
 tL\ge(3N-5x)M=(18t-5s-1)M>0.
\end{equation*}
Thus $L>0$ and
\begin{equation}
 0<u_s=\frac{M}{L}\le\frac{t}{18t-5s-1}.
 \tag{S36}
\end{equation}
The right side is less than $1/4$.  Since $8u-16u^2$ increases on $[0,1/4]$, Eq.~(S33) follows from
\begin{equation}
 \frac{C_s}{I_s}\le
 T_{t,s}:=\frac{(18t-5s-1)^2}{8t(16t-5s-1)}.
 \tag{S37}
\end{equation}

\smsection{Exact node--edge decomposition}{sec:S5}

Set
\begin{equation*}
 r=2t-s,
 \qquad N=6t-2-s=4t+r-2.
\end{equation*}
The $s$ differing coordinates are divided into $a$ entries equal to the first center, $b$ equal to the second, and $c$ equal to neither.  The third class has multiplicity $2^c$.  Put $m=s-c$ and $\delta=|a-b|$.  The larger local distance is
\begin{equation*}
 c+\max(a,b)=\frac{s+c+\delta}{2},
\end{equation*}
so the available outside radius is $(r-c-\delta)/2$.  It follows directly that
\begin{align}
 C_s={}&\sum_{c=0}^{\min(s,r)}\binom{s}{c}2^c
 \sum_{\substack{0\le\delta\le\min(s-c,r-c)\\
                   \delta\equiv s-c\pmod{2}}}
 (2-\delta_{\delta,0})
 \binom{s-c}{(s-c-\delta)/2}
 \Vol_{(r-c-\delta)/2}(N+1).
 \tag{S38}
\end{align}
The double boundary $I_s$ also requires the two local distances to be equal, hence $\delta=0$, and the outside weight to equal $(r-c)/2$.  Therefore
\begin{equation}
 I_s=\sum_{\substack{0\le c\le\min(s,r)\\c\equiv s\pmod{2}}}
 \binom{s}{c}2^c\binom{s-c}{(s-c)/2}w_{(r-c)/2}(N).
 \tag{S39}
\end{equation}

Define the node set
\begin{equation*}
 \mathcal C=\{c:0\le c\le\min(s,r),\ c\equiv s\pmod{2}\},
\end{equation*}
and, for $c\in\mathcal C$,
\begin{equation*}
 m_c=s-c,
 \qquad \ell_c=\frac{r-c}{2},
\end{equation*}
\begin{equation}
 D_c=\binom{s}{c}2^c\binom{m_c}{m_c/2}w_{\ell_c}(N),
 \tag{S40}
\end{equation}
\begin{equation}
 \alpha_c=\sum_{j=0}^{\min(m_c/2,\ell_c)}
 (2-\delta_{j0})
 \frac{\binom{m_c}{m_c/2-j}}{\binom{m_c}{m_c/2}}
 \frac{\Vol_{\ell_c-j}(N+1)}{w_{\ell_c}(N)}.
 \tag{S41}
\end{equation}
These are precisely the even-$\delta$ terms in Eq.~(S38), and Eq.~(S39) becomes $I_s=\sum_cD_c$.

Define the edge set
\begin{equation*}
 \mathcal E=\{e:0\le e\le\min(s,r-1),\ e\not\equiv s\pmod{2}\}.
\end{equation*}
For $e\in\mathcal E$, let
\begin{equation*}
 \rho_e=\frac{r-e-1}{2},
\end{equation*}
\begin{equation}
 B_e=\binom{s}{e}2^e
 \binom{s-e}{(s-e-1)/2}w_{\rho_e}(N),
 \tag{S42}
\end{equation}
\begin{equation}
 \beta_e=2\sum_{j=0}^{\min((s-e-1)/2,\rho_e)}
 \frac{\binom{s-e}{(s-e-1)/2-j}}
      {\binom{s-e}{(s-e-1)/2}}
 \frac{\Vol_{\rho_e-j}(N+1)}{w_{\rho_e}(N)}.
 \tag{S43}
\end{equation}
These are precisely the odd-$\delta$ terms.  Therefore the exact chain decomposition is
\begin{equation}
 I_s=\sum_{c\in\mathcal C}D_c,
 \qquad
 C_s=\sum_{c\in\mathcal C}\alpha_cD_c
 +\sum_{e\in\mathcal E}\beta_eB_e.
 \tag{S44}
\end{equation}

For $e\ge1$, the neighboring nodes are $e-1$ and $e+1$.  Substitution of Eqs.~(S40) and (S42), followed by cancellation of factorials and the adjacent-shell ratio, yields
\begin{equation}
 \frac{B_e^2}{D_{e-1}D_{e+1}}
 =\frac{(e+1)(\rho_e+1)}{3e(N-\rho_e)}
 =:\gamma_e.
 \tag{S45}
\end{equation}
For $e=0$, which occurs only for odd $s$, direct cancellation gives
\begin{equation}
 B_0=\frac{D_1}{s+1}.
 \tag{S46}
\end{equation}
Equations~(S44)--(S46) are the exact combinatorial core of the proof.

\smsection{Uniform local-load bounds}{sec:S6}

For a node $c$, set
\begin{equation*}
 q_c=\frac{\ell_c}{3(N-\ell_c+1)}
 =\frac{r-c}{3(8t+r+c-2)}
 \le q_N:=\frac{r}{3(8t+r-2)}.
\end{equation*}
In Eq.~(S41), the binomial ratio is at most one.  Lemma~\ref{lem:Sshell} gives
\begin{equation*}
 \frac{\Vol_{\ell_c-j}(N+1)}{w_{\ell_c}(N)}
 \le q_c^j\frac{1+3q_c}{1-q_c}.
\end{equation*}
Hence
\begin{align}
 \alpha_c
 &\le\frac{1+3q_N}{1-q_N}
 \left(1+2\sum_{j\ge1}q_N^j\right)\\
 &=\frac{(1+3q_N)(1+q_N)}{(1-q_N)^2}
 =:A(t,r),
 \tag{S47}
\end{align}
where
\begin{equation}
 A(t,r)=
 \frac{3(r+4t-1)(2r+12t-3)}{(r+12t-3)^2}.
 \tag{S48}
\end{equation}

For an ordinary edge $e\ge1$,
\begin{equation*}
 q_e=\frac{\rho_e}{3(N-\rho_e+1)}
 =\frac{r-e-1}{3(8t+r+e-1)}
 \le q_E:=\frac{r-2}{3(8t+r)}.
\end{equation*}
Equation~(S43) similarly gives
\begin{align}
 \beta_e
 &\le2\frac{1+3q_E}{1-q_E}\sum_{j\ge0}q_E^j\\
 &=\frac{2(1+3q_E)}{(1-q_E)^2}
 =:B(t,r),
 \tag{S49}
\end{align}
where
\begin{equation}
 B(t,r)=
 \frac{9(r+8t)(r+4t-1)}{(r+12t+1)^2}.
 \tag{S50}
\end{equation}
Moreover, Eq.~(S45), $(e+1)/e\le2$, $r-e+1\le r$, and $e\ge1$ imply
\begin{equation}
 \gamma_e\le G(t,r):=
 \frac{2r}{3(r+8t-2)}.
 \tag{S51}
\end{equation}
By AM--GM,
\begin{equation}
 \beta_eB_e
 \le B(t,r)\sqrt{G(t,r)}
 \frac{D_{e-1}+D_{e+1}}2.
 \tag{S52}
\end{equation}
Every internal node is adjacent to at most two ordinary edges.  Its total load is therefore at most
\begin{equation}
 A(t,r)+B(t,r)\sqrt{G(t,r)}.
 \tag{S53}
\end{equation}

If $s$ is odd, the special edge $e=0$ contributes directly to the node $c=1$.  Here
\begin{equation*}
 q_0=\frac{r-1}{3(8t+r-1)},
\end{equation*}
Eq.~(S43) gives
\begin{equation}
 \beta_0\le H(t,r):=
 \frac{2(1+3q_0)}{(1-q_0)^2},
 \tag{S54}
\end{equation}
and Eq.~(S46) shows that the special contribution to the $D_1$ load is at most $H(t,r)/(s+1)$.  The same endpoint has only one ordinary neighboring edge, whose charge is at most $B\sqrt{G}/2$.

\smsection{Exact algebraic certificates}{sec:S7}

This section proves the two local-load inequalities used in the main text.  The substitutions below map the relevant integer parameter domains into $x,y\ge0$.  All displayed denominators are products of affine factors that are strictly positive on these domains.  The accompanying exact-arithmetic script reconstructs each rational expression from the displayed formulas, verifies each recorded numerator by cross-multiplication, and checks that every numerator coefficient is strictly positive.  No floating-point sampling, semidefinite optimization, or parameter scan is used.

\subsection{Internal nodes}

Assume $s\ge4$ and $r=2t-s\ge2$.  Set
\begin{equation}
 x=r-2\ge0,
 \qquad y=s-4\ge0,
 \qquad t=\frac{x+y+6}{2}.
 \tag{S55}
\end{equation}
Let $T=T_{t,s}$ and use Eqs.~(S48), (S50), and (S51).  Direct reduction gives
\begin{equation}
 T-A=\frac{P_1(x,y)}
 {4(x+y+6)(7x+6y+35)^2(8x+3y+27)},
 \tag{S56}
\end{equation}
where
\begin{align}
P_1={}&1665x^4+3900x^3y+26556x^3+3244x^2y^2
+45600x^2y+157314x^2\nonumber\\
&+1128xy^3+24676xy^2+176028xy+410400x\nonumber\\
&+144y^4+4272y^3+46960y^2+225552y+398961.
 \tag{S57}
\end{align}
Thus $T-A>0$.  A second reduction gives
\begin{equation}
 (T-A)^2-B^2G=\frac{P_2(x,y)}{D_2(x,y)},
 \tag{S58}
\end{equation}
with
\begin{align}
D_2={}&16(x+y+6)^2(5x+4y+24)
(7x+6y+35)^4\nonumber\\
&\times(7x+6y+39)^4(8x+3y+27)^2.
 \tag{S59}
\end{align}
The coefficient table for $P_2$ appears below.  Every entry is strictly positive, hence $P_2>0$ on $x,y\ge0$.  Equations~(S56)--(S59) imply
\begin{equation}
 A+B\sqrt{G}\le T.
 \tag{S60}
\end{equation}

\subsection{Odd left endpoint}

Assume $s\ge7$ and $r\ge3$ are odd.  Set
\begin{equation}
 x=r-3\ge0,
 \qquad y=s-7\ge0,
 \qquad t=\frac{x+y+10}{2}.
 \tag{S61}
\end{equation}
Define
\begin{equation*}
 E=T-A-\frac{H}{s+1}.
\end{equation*}
Then
\begin{equation}
 E=\frac{P_3(x,y)}
 {4(y+8)(x+y+10)(7x+6y+60)^2
 (7x+6y+62)^2(8x+3y+44)},
 \tag{S62}
\end{equation}
and
\begin{equation}
 E^2-\frac{1}{4}B^2G=\frac{P_4(x,y)}{D_4(x,y)},
 \tag{S63}
\end{equation}
where
\begin{align}
D_4={}&16(y+8)^2(x+y+10)^2(5x+4y+41)
(7x+6y+60)^4\nonumber\\
&\times(7x+6y+62)^4(7x+6y+64)^4
(8x+3y+44)^2.
 \tag{S64}
\end{align}
The coefficient tables below show that both $P_3$ and $P_4$ have strictly positive coefficients.  Therefore
\begin{equation}
 A+\frac{H}{s+1}+\frac{1}{2}B\sqrt{G}\le T.
 \tag{S65}
\end{equation}

For compactness, the tables encode a polynomial as
\begin{equation*}
 P_k(x,y)=\sum_i x^ip_i(y),
 \qquad
 p_i(y)=\sum_jp_{i,j}y^j,
\end{equation*}
listing the coefficient vector $(p_{i,0},p_{i,1},\ldots)$.  This is also a directly checkable positive-coefficient certificate.

\subsubsection*{\texorpdfstring{Coefficient vectors for $P_2$}{Coefficient vectors for P2}}
\begingroup\small\raggedright\sloppy
Write $P_2(x,y)=\sum_i x^i p_i(y)$ and $p_i(y)=\sum_j p_{i,j}y^j$. The vectors below list $(p_{i,0},p_{i,1},\ldots)$.\par\medskip
\noindent $i=0:\quad (1062899891563722264,\allowbreak 2469168164232512964,\allowbreak 2617958587262949216,\allowbreak 1679939647381932768,\allowbreak 728729213700228480,\allowbreak 225836803667715264,\allowbreak 51472473830461440,\allowbreak 8739756654686208,\allowbreak 1105900835014656,\allowbreak 103024844172288,\allowbreak 6869505024000,\allowbreak 310500532224,\allowbreak 8527970304,\allowbreak 107495424)$\par\smallskip
\noindent $i=1:\quad (2483601030292829733,\allowbreak 5348862818014093560,\allowbreak 5228089967125216152,\allowbreak 3069982755406067040,\allowbreak 1207197497524602672,\allowbreak 335094299849035008,\allowbreak 67358347527435264,\allowbreak 9882898470973440,\allowbreak 1050712475680512,\allowbreak 78957039015936,\allowbreak 3981434923008,\allowbreak 120974155776,\allowbreak 1675137024)$\par\smallskip
\noindent $i=2:\quad (2699270624171263428,\allowbreak 5315130738588826584,\allowbreak 4726297260855550008,\allowbreak 2505565019113058736,\allowbreak 879995803598384256,\allowbreak 215022785157403200,\allowbreak 37302342282565632,\allowbreak 4594901389187328,\allowbreak 393883117231104,\allowbreak 22379600471040,\allowbreak 758578784256,\allowbreak 11621449728)$\par\smallskip
\noindent $i=3:\quad (1834738746247840914,\allowbreak 3240360026253725256,\allowbreak 2573963814247512480,\allowbreak 1209395775189052512,\allowbreak 371839443065265456,\allowbreak 78108307633298496,\allowbreak 11345832816802368,\allowbreak 1124854493303808,\allowbreak 72823913872128,\allowbreak 2779498810368,\allowbreak 47486352384)$\par\smallskip
\noindent $i=4:\quad (887227815107452644,\allowbreak 1371763991511722556,\allowbreak 950105457220794336,\allowbreak 385855476109932960,\allowbreak 101014333852057056,\allowbreak 17642381790204544,\allowbreak 2052275085484800,\allowbreak 153133135593984,\allowbreak 6644148926976,\allowbreak 127624522752)$\par\smallskip
\noindent $i=5:\quad (328215861855860283,\allowbreak 432772050040076304,\allowbreak 253756134072067824,\allowbreak 86181149159291424,\allowbreak 18484133912260400,\allowbreak 2555733903572864,\allowbreak 221830101742464,\allowbreak 11023894858752,\allowbreak 239669618688)$\par\smallskip
\noindent $i=6:\quad (96382747446713640,\allowbreak 105901522408113840,\allowbreak 50918050161104400,\allowbreak 13878628903412160,\allowbreak 2311438768758976,\allowbreak 234519725011008,\allowbreak 13375266297600,\allowbreak 329629881600)$\par\smallskip
\noindent $i=7:\quad (22615275375926892,\allowbreak 20335395348931248,\allowbreak 7766982233572224,\allowbreak 1617040813447968,\allowbreak 193743412123728,\allowbreak 12656760156992,\allowbreak 351380071488)$\par\smallskip
\noindent $i=8:\quad (4169496912491520,\allowbreak 3013715284716636,\allowbreak 882819741555072,\allowbreak 131600194069632,\allowbreak 10023043258720,\allowbreak 312888506944)$\par\smallskip
\noindent $i=9:\quad (585067476490587,\allowbreak 330466355568504,\allowbreak 70401947613624,\allowbreak 6734064619776,\allowbreak 245260577312)$\par\smallskip
\noindent $i=10:\quad (59768040946068,\allowbreak 25001528421240,\allowbreak 3484930590456,\allowbreak 162367507344)$\par\smallskip
\noindent $i=11:\quad (4163896592226,\allowbreak 1155490422408,\allowbreak 79804107936)$\par\smallskip
\noindent $i=12:\quad (176187376260,\allowbreak 24444779940)$\par\smallskip
\noindent $i=13:\quad (3408279525)$\par\smallskip
\endgroup
Every displayed coefficient is strictly positive, so $P_2(x,y)>0$ for $x,y\ge0$.

\par\medskip
\subsubsection*{\texorpdfstring{Coefficient vectors for $P_3$}{Coefficient vectors for P3}}
\begingroup\small\raggedright\sloppy
Write $P_3(x,y)=\sum_i x^i p_i(y)$ and $p_i(y)=\sum_j p_{i,j}y^j$. The vectors below list $(p_{i,0},p_{i,1},\ldots)$.\par\medskip
\noindent $i=0:\quad (45086423040,\allowbreak 34588652544,\allowbreak 10992354624,\allowbreak 1891841760,\allowbreak 191452320,\allowbreak 11433888,\allowbreak 374112,\allowbreak 5184)$\par\smallskip
\noindent $i=1:\quad (41722881024,\allowbreak 27878107776,\allowbreak 7478076288,\allowbreak 1040737584,\allowbreak 79726656,\allowbreak 3199824,\allowbreak 52704)$\par\smallskip
\noindent $i=2:\quad (15701105664,\allowbreak 8964529584,\allowbreak 1960785200,\allowbreak 207892792,\allowbreak 10761768,\allowbreak 218592)$\par\smallskip
\noindent $i=3:\quad (3081509184,\allowbreak 1461818448,\allowbreak 246320336,\allowbreak 17777356,\allowbreak 468168)$\par\smallskip
\noindent $i=4:\quad (333546336,\allowbreak 126076368,\allowbreak 14724440,\allowbreak 546496)$\par\smallskip
\noindent $i=5:\quad (18928560,\allowbreak 5329380,\allowbreak 330960)$\par\smallskip
\noindent $i=6:\quad (441000,\allowbreak 81585)$\par\smallskip
\endgroup
Every displayed coefficient is strictly positive, so $P_3(x,y)>0$ for $x,y\ge0$.

\par\medskip
\subsubsection*{\texorpdfstring{Coefficient vectors for $P_4$}{Coefficient vectors for P4}}
\begingroup\small\raggedright\sloppy
Write $P_4(x,y)=\sum_i x^i p_i(y)$ and $p_i(y)=\sum_j p_{i,j}y^j$. The vectors below list $(p_{i,0},p_{i,1},\ldots)$.\par\medskip
\noindent $i=0:\quad (22296658450431682375424409600,\allowbreak 410157638920461592191691653120,\allowbreak 675588887321446906318276263936,\allowbreak 542001476215763101307300216832,\allowbreak 274559512741685245931467309056,\allowbreak 97917261306259441726115020800,\allowbreak 26064826696689045131603607552,\allowbreak 5363497048845964563821887488,\allowbreak 872111099897430344954118144,\allowbreak 113579291280223527264681984,\allowbreak 11935560964868041332400128,\allowbreak 1014442016263899402977280,\allowbreak 69570476222305031184384,\allowbreak 3821274181511236878336,\allowbreak 165828126171760386048,\allowbreak 5560688374147104768,\allowbreak 139034575005622272,\allowbreak 2440941160955904,\allowbreak 26852786896896,\allowbreak 139314069504)$\par\smallskip
\noindent $i=1:\quad (177183467552492013151290654720,\allowbreak 1123234415014254579812602478592,\allowbreak 1590599915627187339688245460992,\allowbreak 1160466934210708179293262839808,\allowbreak 541783722011397109285313839104,\allowbreak 178585054571244997535289311232,\allowbreak 43877367379338359990082600960,\allowbreak 8302157745749098733111869440,\allowbreak 1234318365368459135317573632,\allowbreak 145896813020240108126797824,\allowbreak 13784117440874168059232256,\allowbreak 1040745289032371681353728,\allowbreak 62433809716859582717952,\allowbreak 2939185744181806546944,\allowbreak 106302431997586980864,\allowbreak 2852396964600569856,\allowbreak 53507904382181376,\allowbreak 626541808582656,\allowbreak 3448023220224)$\par\smallskip
\noindent $i=2:\quad (341774760347314254784355106816,\allowbreak 1409685416858576307936710098944,\allowbreak 1747551816458261388296262254592,\allowbreak 1160386676238712798178371436544,\allowbreak 497898873445040584050916786176,\allowbreak 151031454167316364577628880896,\allowbreak 34061956025475800673065336832,\allowbreak 5886665358229425629743349760,\allowbreak 793779113048395358374834176,\allowbreak 84316491054908283007107072,\allowbreak 7074521977939081293293568,\allowbreak 467147649142276905074688,\allowbreak 24015808704198002982912,\allowbreak 942182200195704446976,\allowbreak 27266297394263574528,\allowbreak 548902722136891392,\allowbreak 6867467223293952,\allowbreak 40226937569280)$\par\smallskip
\noindent $i=3:\quad (337840891431646826697131556864,\allowbreak 1075131986677789535950111506432,\allowbreak 1183445234962104246251190484992,\allowbreak 715910672711949310943074713600,\allowbreak 281564192350365556390871728128,\allowbreak 78238294538206137419244060672,\allowbreak 16098193846980880072409014272,\allowbreak 2521668317858630005618864128,\allowbreak 305456965300784294870581248,\allowbreak 28808769803132770089394176,\allowbreak 2113767165245216273602560,\allowbreak 119610101333663534333952,\allowbreak 5124446311671517151232,\allowbreak 160871198049112117248,\allowbreak 3493004781939664896,\allowbreak 46902320681951232,\allowbreak 293575538958336)$\par\smallskip
\noindent $i=4:\quad (211189970432380384154952400896,\allowbreak 558751133088882105432603623424,\allowbreak 551524056204383488650050371584,\allowbreak 304027226031651206758493700096,\allowbreak 109248260458541165174518851584,\allowbreak 27660696720231671699877218304,\allowbreak 5155386439526730065989433856,\allowbreak 725235050877727293076974592,\allowbreak 77993598980760493608036864,\allowbreak 6432578617030749609338880,\allowbreak 404485617200658180148224,\allowbreak 19076507918789472552960,\allowbreak 654080220287019518976,\allowbreak 15408395823247294464,\allowbreak 223189113987495936,\allowbreak 1499556166262784)$\par\smallskip
\noindent $i=5:\quad (92028643489687399894049882112,\allowbreak 210407466878218873564277243904,\allowbreak 187370695615351715785520283648,\allowbreak 94023264645023155574631579648,\allowbreak 30738791012896431933004187648,\allowbreak 7045485812695686684117023744,\allowbreak 1179057377402086341231675392,\allowbreak 147259084135971389329282560,\allowbreak 13850751486984923109846528,\allowbreak 979196542336752166232064,\allowbreak 51329923948778555990016,\allowbreak 1937862923796603712512,\allowbreak 49871886321457529856,\allowbreak 783938748234240000,\allowbreak 5683238980319232)$\par\smallskip
\noindent $i=6:\quad (29507998884985217796025614336,\allowbreak 59568122502554579185756078080,\allowbreak 47997321458506070237998006272,\allowbreak 21868490153776169033267189760,\allowbreak 6468050767971710927001201664,\allowbreak 1331063911324687784428617216,\allowbreak 197804660178732645853010304,\allowbreak 21613585070092124879105024,\allowbreak 1743171817342523946587136,\allowbreak 102772531374311897266176,\allowbreak 4314107045266889225472,\allowbreak 122291426181108888576,\allowbreak 2100752249102668800,\allowbreak 16532523459477504)$\par\smallskip
\noindent $i=7:\quad (7191033294214938982245728256,\allowbreak 12972998621613933031887175680,\allowbreak 9460638120546353916299759616,\allowbreak 3896859972853355422425237504,\allowbreak 1034966308889871720875565568,\allowbreak 189214627269113819113553152,\allowbreak 24613347184074407385852544,\allowbreak 2307426161021331013509120,\allowbreak 155281697428031195226624,\allowbreak 7334943408586640194560,\allowbreak 231299941070850308352,\allowbreak 4378491590785827840,\allowbreak 37672592025811968)$\par\smallskip
\noindent $i=8:\quad (1358144493076377780529397760,\allowbreak 2203365797220439503380201472,\allowbreak 1451195819529501734711705856,\allowbreak 536981259733488398315515392,\allowbreak 126803623675867000086717696,\allowbreak 20311133654321344712577408,\allowbreak 2268830668601334614929632,\allowbreak 177621934189201352352960,\allowbreak 9584204930258565682080,\allowbreak 340325620130580523776,\allowbreak 7171115161797040896,\allowbreak 68030312894346240)$\par\smallskip
\noindent $i=9:\quad (200916857196005785682706432,\allowbreak 293798078392154519057651712,\allowbreak 173990216312929515571292928,\allowbreak 57328800193241989651593216,\allowbreak 11879157042154082055804288,\allowbreak 1636264908661744858187328,\allowbreak 152829062053367238533568,\allowbreak 9603956457436743820896,\allowbreak 389939222986343889312,\allowbreak 9260280495038674944,\allowbreak 97862365073053440)$\par\smallskip
\noindent $i=10:\quad (23363781591379450215665664,\allowbreak 30772898697395086374159360,\allowbreak 16273331841919341876906240,\allowbreak 4717646784583334895549312,\allowbreak 842591175691416894971200,\allowbreak 97237695110928968645280,\allowbreak 7302531737356980885096,\allowbreak 345807439682897531360,\allowbreak 9401791338348314976,\allowbreak 112099914346210560)$\par\smallskip
\noindent $i=11:\quad (2129144281207630020182016,\allowbreak 2517780476865751351865856,\allowbreak 1177246173720749636644224,\allowbreak 295368457509512763360768,\allowbreak 44345293422877116392800,\allowbreak 4126028904361690632880,\allowbreak 234025302851638792648,\allowbreak 7434758839028930976,\allowbreak 101647425125750112)$\par\smallskip
\noindent $i=12:\quad (150501793854031257996288,\allowbreak 158881652823742786199808,\allowbreak 64787266883629994497248,\allowbreak 13745355187486644914976,\allowbreak 1671571523630135691368,\allowbreak 117897664129438924416,\allowbreak 4499582985989528472,\allowbreak 72058836678668896)$\par\smallskip
\noindent $i=13:\quad (8091871303696733500416,\allowbreak 7562821098835226335104,\allowbreak 2638012692277866855168,\allowbreak 457285884531248824128,\allowbreak 42433511627127541088,\allowbreak 2024122309245733424,\allowbreak 39098278910083736)$\par\smallskip
\noindent $i=14:\quad (319995571283490703104,\allowbreak 261751445465855615424,\allowbreak 75938949579568882128,\allowbreak 10176455421913446288,\allowbreak 644831894973773304,\allowbreak 15680993670479424)$\par\smallskip
\noindent $i=15:\quad (8779829899449659904,\allowbreak 6187928670639606240,\allowbreak 1429548575773135608,\allowbreak 133604633235369240,\allowbreak 4380936577907904)$\par\smallskip
\noindent $i=16:\quad (149364078810166080,\allowbreak 88768257715614960,\allowbreak 15145375888108305,\allowbreak 761332578250860)$\par\smallskip
\noindent $i=17:\quad (1187170835054400,\allowbreak 576962037363600,\allowbreak 61975790230725)$\par\smallskip
\endgroup
Every displayed coefficient is strictly positive, so $P_4(x,y)>0$ for $x,y\ge0$.

\par\medskip

\smsection{Low-support families, terminal family, and starting values}{sec:S8}

The uniform internal-node estimate covers $s\ge4$, $r\ge2$.  The odd endpoint estimate covers the odd subregion $s\ge7$, $r\ge3$.  The remaining families are evaluated below.

\subsection{\texorpdfstring{The family $s=2$}{The family s=2}}

Direct classification of the two differing coordinates gives
\begin{align}
 C_2&=2\Vol_{t-1}(6t-3)+14\Vol_{t-2}(6t-3),\tag{S66}\\
 I_2&=2w_{t-1}(6t-4)+4w_{t-2}(6t-4).
 \tag{S67}
\end{align}
Set
\begin{equation*}
 a=\frac{t-1}{3(5t-2)}.
\end{equation*}
Applying Lemma~\ref{lem:Sshell} to Eqs.~(S66)--(S67), together with the length recursion, gives
\begin{equation}
 \frac{C_2}{I_2}\le
 F_2(a):=\frac{(1+3a)(1+7a)}{(1-a)(1+2a)}.
 \tag{S68}
\end{equation}
With $x=t-1\ge0$,
\begin{equation}
 T_{t,2}-F_2(a)=
 \frac{26424x^4+37764x^3+21526x^2+6075x+729}
 {8t(14t-5)(16t-11)(17t-8)}>0.
 \tag{S69}
\end{equation}

\subsection{\texorpdfstring{The family $s=3$}{The family s=3}}

Here
\begin{align}
 C_3&=18\Vol_{t-2}(6t-4)+46\Vol_{t-3}(6t-4),\tag{S70}\\
 I_3&=12w_{t-2}(6t-5)+8w_{t-3}(6t-5).
 \tag{S71}
\end{align}
Let
\begin{equation*}
 a=\frac{t-2}{3(5t-2)}.
\end{equation*}
The shell-tail bound gives
\begin{equation}
 \frac{C_3}{I_3}\le
 F_3(a):=\frac{1+3a}{1-a}\frac{18+46a}{12+8a}.
 \tag{S72}
\end{equation}
With $x=t-2\ge0$,
\begin{equation}
 T_{t,3}-F_3(a)=
 \frac{3897x^4+16032x^3+23828x^2+15024x+3456}
 {32t(t-1)(7t-2)(47t-22)}>0.
 \tag{S73}
\end{equation}

\subsection{\texorpdfstring{The family $s=5$}{The family s=5}}

For $t\ge5$,
\begin{align}
 C_5={}&80\Vol_{t-3}(6t-6)+490\Vol_{t-4}(6t-6)
 +454\Vol_{t-5}(6t-6),\tag{S74}\\
 I_5={}&60w_{t-3}(6t-7)+160w_{t-4}(6t-7)
 +32w_{t-5}(6t-7).
 \tag{S75}
\end{align}
Let
\begin{equation*}
 a=\frac{t-3}{3(5t-3)}.
\end{equation*}
Dropping the last positive term of $I_5$ and bounding the numerator tails gives
\begin{equation}
 \frac{C_5}{I_5}\le
 F_5(a):=\frac{1+3a}{1-a}
 \frac{80+490a+454a^2}{60+160a}.
 \tag{S76}
\end{equation}
With $x=t-5\ge0$,
\begin{align}
 T_{t,5}-F_5(a)
 ={}&\frac{131979x^5+2477928x^4+18612008x^3
 +69931704x^2}{20t(5t-3)(7t-3)(8t-13)(53t-51)}\nonumber\\
 &+\frac{131509536x+99097600}
 {20t(5t-3)(7t-3)(8t-13)(53t-51)}>0.
 \tag{S77}
\end{align}
The omitted values are
\begin{equation*}
 t=3:\quad \frac{C_5}{I_5}=\frac{4}{3}<\frac{49}{33}=T_{3,5},
\end{equation*}
\begin{equation*}
 t=4:\quad \frac{C_5}{I_5}=\frac{489}{322}<\frac{529}{304}=T_{4,5}.
\end{equation*}

\subsection{\texorpdfstring{The terminal family $r=1$}{The terminal family r=1}}

When $s=2t-1$, the chain consists of the node $c=1$ and the special edge $e=0$.  Hence
\begin{equation}
 \frac{C_{2t-1}}{I_{2t-1}}=\frac{t+1}{t}.
 \tag{S78}
\end{equation}
Furthermore,
\begin{equation}
 T_{t,2t-1}-\frac{t+1}{t}
 =\frac{t^2-t-1}{t(3t+2)}>0
 \qquad(t\ge2).
 \tag{S79}
\end{equation}
For clarity, we partition the parameter range as follows:
\[
\begin{array}{c|c}
\text{parameter region}&\text{estimate used}\\ \hline
s=2&\text{Eq.~(S69)}\\
s=3&\text{Eq.~(S73)}\\
s=5&\text{Eq.~(S77) and the two explicit values}\\
r=1,\ s\ge7&\text{Eq.~(S79)}\\
s\ge4\ \text{even},\ r\ge2&\text{Eq.~(S60)}\\
s\ge7\ \text{odd},\ r\ge3&\text{Eqs.~(S60) and (S65)}
\end{array}
\]
Indeed, $r=2t-s$ has the same parity as $s$.  Thus an even $s$ outside the low-support cases has $r\ge2$, while an odd $s\ge7$ has either $r=1$ or $r\ge3$.  Hence the table is exhaustive for $2\le s\le2t-1$.  Therefore
\begin{equation}
 R_{6t-1,t}(2)\ge R_{6t-1,t}(3)\ge\cdots\ge R_{6t-1,t}(2t).
 \tag{S80}
\end{equation}

\subsection{Starting values}

For $s=1$,
\begin{equation*}
 L_1=w_t(6t-2),\qquad C_1=4\Vol_{t-1}(6t-2).
\end{equation*}
Using Lemma~\ref{lem:Sshell} and the length recursion gives
\begin{equation}
 R_{6t-1,t}(1)
 \le\frac{12t(6t-1)}{(14t-3)^2}<\frac{1}{2}.
 \tag{S81}
\end{equation}

For $s=2$, put
\begin{equation*}
 q=\frac{t}{3(5t-2)},
 \qquad W=w_t(6t-3).
\end{equation*}
The shift identity gives
\begin{equation*}
 L_2=W(1-q).
\end{equation*}
Two applications of the length recursion and Lemma~\ref{lem:Sshell} yield
\begin{equation*}
 \Vol_t(6t-1)\le W\frac{(1+3q)^2}{1-q},
\end{equation*}
while Eq.~(S66) gives
\begin{equation*}
 C_2\le W\frac{2q+14q^2}{1-q}.
\end{equation*}
Therefore
\begin{equation}
 R_{6t-1,t}(2)
 \le\frac{9t(3t-1)^2(11t-3)}{(7t-3)^4}<1.
 \tag{S82}
\end{equation}
To verify the final inequality, set $x=t-2\ge0$. Then
\begin{equation}
 1-\frac{9t(3t-1)^2(11t-3)}{(7t-3)^4}
 =\frac{1510x^4+8801x^3+18951x^2+17783x+6091}
 {(7x+11)^4}>0.
 \tag{S83}
\end{equation}
For $t=1$, direct evaluation gives $R_{5,1}(1)=4/9$ and $R_{5,1}(2)=1/2$.

Combining Eqs.~(S80)--(S83) gives
\begin{equation}
 R_{6t-1,t}(s)<1,
 \qquad 1\le s\le2t.
 \tag{S84}
\end{equation}
For $t=0$, the bound is the trivial dimension estimate $K\le2^n$.  For $t\ge1$, length monotonicity, the Rains shadow bound $t\le\lfloor(n+1)/6\rfloor$, and the Li--Xing reduction prove the theorem in the main text.